# Venus: Key to understanding the evolution of terrestrial planets

A response to ESA's Call for White Papers for the Definition of Science Themes for L2/L3 Missions in the ESA Science Programme.

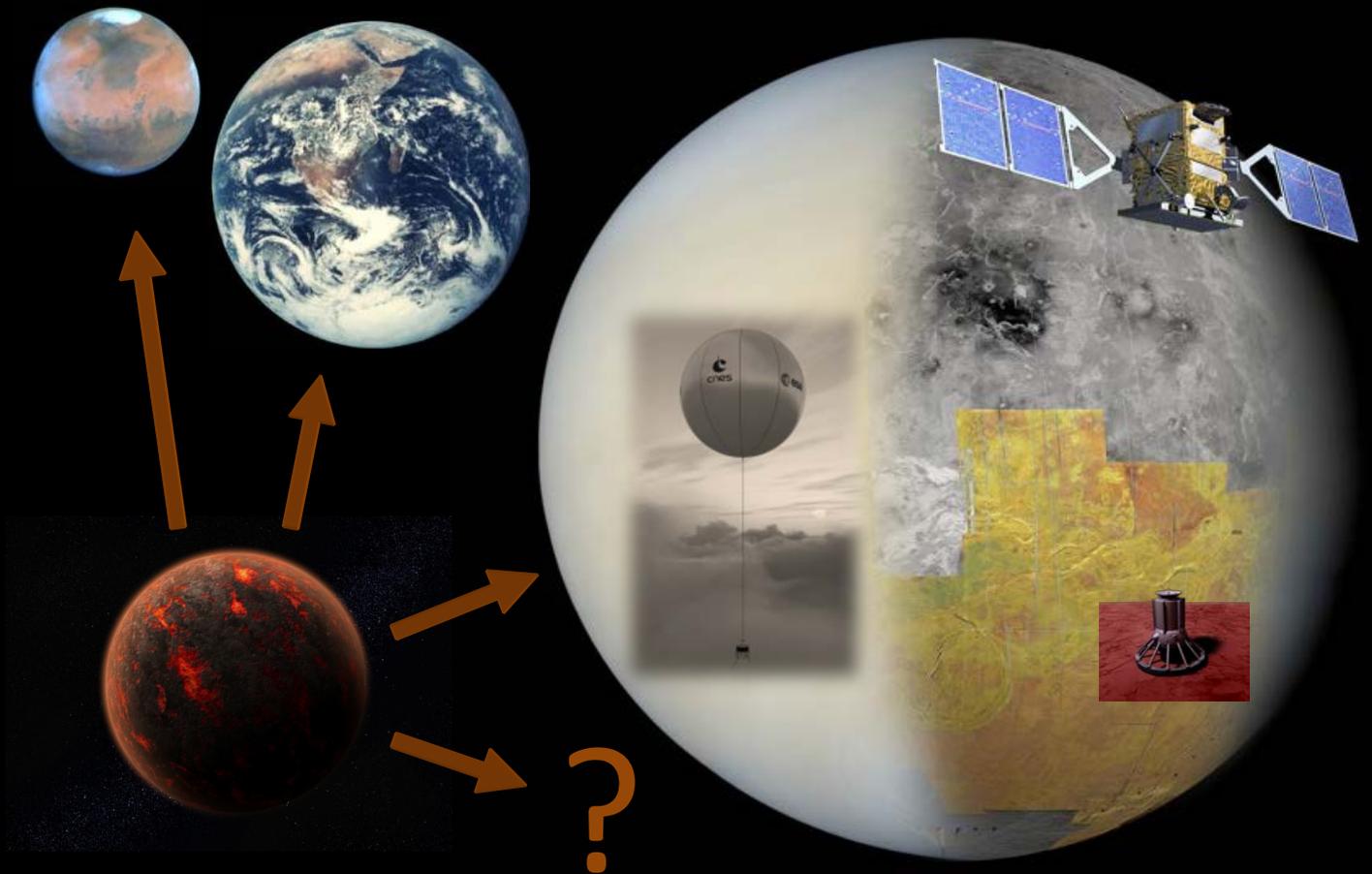

Spokesperson: Colin Wilson

Atmospheric, Oceanic and Planetary Physics,
Clarendon Laboratory,
University of Oxford, UK.
E-mail: wilson@atm.ox.ac.uk

# Executive summary

In this White Paper, we advocate L2/L3 science themes of understanding the diversity and evolution of habitable planets, and emphasize the importance of Venus to these science themes.

Why are the terrestrial planets so different from each other? Venus should be the most Earth-like of all our planetary neighbours. Its size, bulk composition and distance from the Sun are very similar to those of the Earth. Its original atmosphere was probably similar to that of early Earth, with large atmospheric abundances of carbon dioxide and water. Furthermore, the young sun's fainter output may have permitted a liquid water ocean on the surface. While on Earth a moderate climate ensued, Venus experienced runaway greenhouse warming, which led to its current hostile climate. How and why did it all go wrong for Venus? What lessons can we learn about the life story of terrestrial planets/exoplanets in general, whether in our solar system or in others?

ESA's Venus Express mission has proved tremendously successful, answering many questions about Earth's sibling planet and establishing European leadership in Venus research. However, further understanding of Venus and its history requires several further lines of investigation. Entry into the atmosphere will be required in order to measure noble gas isotopic signatures of past history and to understand the role of clouds in the climate balance. Radar mapping at metre-scale spatial resolution, and surface height change detection at centimetre scale, would enable detection of current volcanic and Aeolian activity, and would revolutionise comparative geology between the terrestrial planets. The tessera highlands of Venus are thought to be the oldest terrain type found on Venus but have not yet been visited by spacecraft; a lander in these regions would measure surface composition to provide clues as to the earliest geologic record available on Venus.

Individually, these investigations could be carried out by separate low cost missions, and there is ample scope for international collaboration as individual payload or mission elements could be provided or even separately launched by different space agencies. However, there is strong synergy to be achieved by having all of these elements operating at the same time at Venus in close co-ordination. An orbiter permits much higher data return from balloons or landing probes by acting as data relay, and also provides positioning and context imaging for in situ measurements, as has been demonstrated at Mars and Titan. More broadly speaking, the combination of geological data from surface and radar elements with new geochemical constraints from isotopic ratio measurements and in situ atmospheric data would reveal the evolution of Venus and its climate, with relevance to terrestrial planets everywhere.

To address these themes **we propose a strawman mission based on a combination of an in situ balloon platform, a radar-equipped orbiter, and (optionally) a descent probe.** These mission elements are modelled on the 2010 EVE M3 mission proposal, on, the 2010 EnVision M3 proposal (which proposes re-use of ESA's GMES Sentinel-1 radar technology), and on Russia's Venera-D entry probe, respectively. The science themes described in the present white paper are phrased specifically in terms of Venus, but could equally be included in broader science themes such as comparative planetology or solar system evolution, both of which are central to our understanding of our own planet and of the diversity of exoplanets being discovered.

Written by Colin Wilson (Oxford University, UK) with contributions from K. Baines (U. Wisconsin & JPL, USA), E. Chassefière (Univ. Paris-Sud, France), R. Ghail (Imperial College, UK), R. Ghent (U. Toronto, Canada), J. Helbert (DLR, Germany), and the EVE and Envision Science Teams.

# 1 Introduction

## 1.1 The importance of Venus for comparative planetology

One of the central goals of planetary research is to find our place in the Universe. Investigation the evolution of solar systems, planets, and of life itself, is at the heart of this quest. The three terrestrial planets in our solar system - Earth, Mars, and Venus - show a wide range of evolutionary pathways, and so represent a "key" to our understanding of planets and exoplanets.

Earth and Venus were born as twins – formed at around the same time, with apparently similar bulk composition and the same size. However, they have evolved very differently: the enormous contrast between these planets today challenge our understanding of how terrestrial planets work. The atmosphere is surprising in many ways – its 400 km/h winds on a slowly rotating planet; its enormous surface temperature, even though it absorbs *less* sunlight than does the Earth; its extreme aridity, with sulphuric acid as its main condensable species instead of water. The solid planet, too is mysterious: its apparent lack of geodynamo and plate tectonics, the uncertainty of its current volcanic state, the apparent young age of much of its surface. How and why does a planet so similar to Earth end up so different?

In an era where we will soon have detected hundreds and then thousands of Earth-sized exoplanets, planetary science must seek to characterise these planets and to explain the diverse outcomes which may befall them. Are these exoplanets habitable? Many efforts have been made to define the edges of the 'habitable zone', i.e. the range of distances from a parent star at which a planet can sustain liquid water on its surface. The inner edge of the habitable zone has been estimated to lie anywhere from 0.5 to 0.99 AU – this latter figure, from Kopparapu et al., 2013, should be a cause of concern for us Earth-dwellers! Detailed study of Venus is indispensable if we are to understand what processes determine the inner edge of the habitable zone.

The habitable zone's boundaries will evolve over a planet's lifetime due to the evolution of the star's output as well as changes in the planet and its atmosphere. There are only three terrestrial planets at which we can study geophysical and evolutionary processes: Venus, Earth, and Mars. Exploration of the latter two is firmly established, while in contrast there are no Venus missions currently planned after Venus Express.

## 1.2 Context: The state of Venus science after Venus Express

Venus Express has been a tremendously successful mission. Since its arrival at Venus in April 2006 it has made a wealth of discoveries relating to the atmosphere at all altitudes from the surface up to the exosphere. It has mapped cloud motions to reveal wind velocities at different altitudes; it has measured the spatial distributions of key chemical species, and discovered new ones; it has measured the rate at which oxygen and hydrogen are being lost to space; it has found signs of frequent lightning; it has found signs of recent lava flows on the surface using surface mapping at 1 μm wavelength.

In short Venus Express has provided much-needed data for atmospheric dynamics, chemistry, and radiative transfer, and for understanding of its ionosphere & induced magnetosphere. These new data are invaluable for constraining models of how Venus works today. However, the history of Venus still remains enigmatic. In this paper we propose a new set of investigations that focus on understanding the evolution of Venus through a combination of surface and atmospheric investigations.

Thanks to Venus Express, Europe is at the forefront of Venus research. Research groups across Europe have participated in the construction of and analysis from the scientific payload; dozens of researchers have completed doctoral theses based on Venus Express research. Europe is thus well-placed to lead a future Venus mission. ESA can now build on this position, capitalising on investments made in Earth Observation programme and advanced satellite technologies, to address fundamental questions about the evolution of terrestrial planets and the appearance of life.

# 2 Science Themes

## 2.1 Geology (Interior, tectonism, volcanism, geomorphology, mineralogy)

The major unknowns in Venus geological science are associated with its resurfacing history and establishing whether it is currently geologically active.

**Resurfacing history**

The age of Venus' surface is poorly known. Unlike Mercury, the Moon, and Mars, Venus has a thick atmosphere that represents a powerful filter to small impactors. As a result, its crater population is limited to a few large craters; there are very few craters with diameters <20 km, and there are fewer than 1000 craters in total. The observed crater population offers poor constraints on surface ages, allowing a number of different production and resurfacing scenarios. These include catastrophic global lithospheric overturn which take place every 500 to 700 My [Turcotte et al., 1999], equilibrium resurfacing models more similar to those found on Earth [Stofan et al. 2005], as well as many models in between.

The community has used the existing Magellan radar data to attempt to resolve these fundamental conflicts by applying mapping techniques to establish stratigraphic relationships among surface units and structures. NASA's Magellan orbiter, launched in 1989, obtained global radar maps with a spatial resolution of 100 – 200 m. An important limitation to using radar images for geological mapping is that geological mapping requires the ability to identify distinct rock units, whose formation represent geological processes (e.g., distinct lava flows or sedimentary units). Magellan imagery provides the opportunity to identify some units; for example, it is possible to map lava flow boundaries to high precision in some terrains. But in many, many other cases, the materials being mapped have been affected by later tectonic structures; moreover, the highly deformed tessera terrain, thought to be the oldest on Venus, is characterized by overlapping structures whose relationships are ambiguous in currently available observations. Different methods for accommodating this complication have led to widely divergent mapping styles, which in turn have resulted in a range of surface evolution models, mirroring the range of interior evolution models (see e.g. review by Guest & Stofan, 1999). Many remaining debates over, for instance, the sequence and relative timing of tectonic deformation in the complex tesserae cannot be resolved using currently available radar data, because of the limitations represented by the spatial resolution and single polarization of those data.

**Current geological activity**

Venus is thought to have similar internal heat production to Earth, but it is not clear how the internal heat is lost to space. Is heat lost solely by crustal conduction, or does volcanic activity play an important role? Is the loss rate sufficient to maintain an equilibrium or is heat building up in the interior, potentially leading to an episodic resurfacing scenario? Understanding how Venus loses its internal heat is important for understanding both Earth's earliest history and for understanding those exoplanets larger than Earth, both of which share its problem of a buoyant lithosphere; these mechanisms at work may profoundly affect the atmosphere and climate, and prove catastrophic for life.

There are some hints of recent geological activity particularly from Venus Express data (Smrekar et al., 2010, Marcq et al., 2012), but these analyses are indirect. A new radar dataset would enable not only better understanding of current surface weathering and alteration processes, which is needed in order to calculate ages for geologically recent changes such as lava flows and dune movements, but would also enable direct searching for surface change.

**Case for next-generation radar:**

Because of the extreme surface conditions and opaque clouds on Venus, geological investigation requires orbital remote sensing, with techniques including interferometric synthetic aperture radar (InSAR), gravimetry, altimetry and infrared observation using nightside infrared windows. The value of radar mapping at Venus was demonstrated by NASA's Magellan orbiter, launched in 1989, which obtained global radar

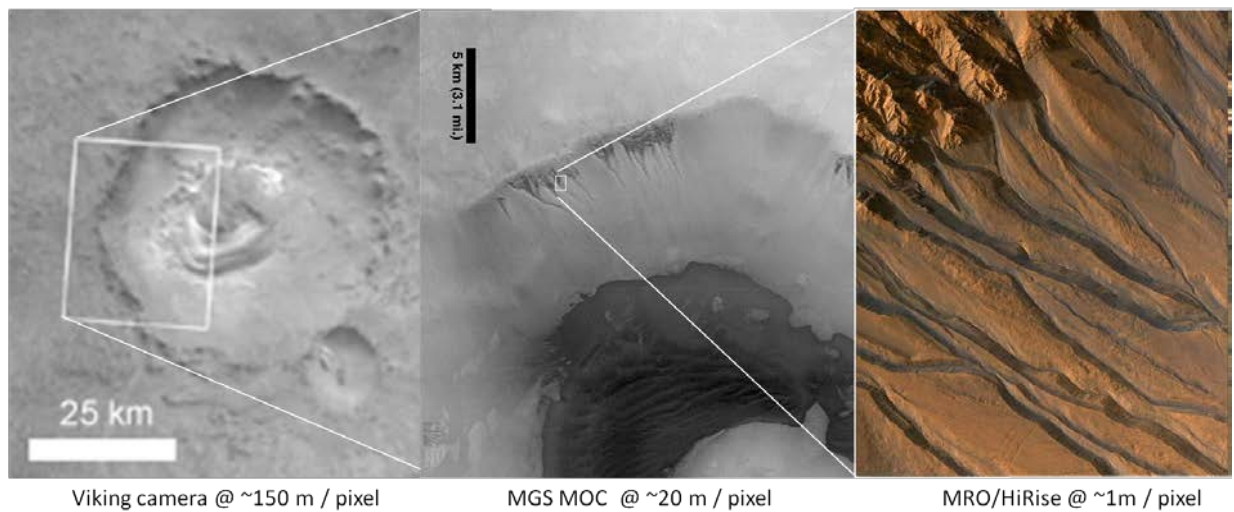

Viking camera @ ~150 m / pixel     MGS MOC @ ~20 m / pixel     MRO/HiRise @ ~1m / pixel

*Images of the same region of Mars illustrate the revolutions in understanding which are enabled by increasing spatial resolution by an order of magnitude, particular for surface processes. Quoted pixel resolution is that of the displayed image used rather than the full resolution of the original. Image credits: NASA/JPL.*

maps with a horizontal resolution of 100 – 200 m and altimetry with a vertical resolution of 100 m. Advances in technology, data acquisition and processing, and satellite control and tracking, mean that the spatial resolutions in the 1 – 10 m range are now possible.

This high-resolution radar mapping of Venus would revolutionise geological understanding. Generations of Mars orbital imagery have seen successive order-of-magnitude improvements, as illustrated below. As imagers progressed from the 50 m resolution of Viking towards the 5m resolution of MGS/MOC and the higher resolutions of MEx/HRSC and MRO/HiRise, our conception of Mars as a frozen, inactive planet was followed by hypotheses that geologically recent flow had occurred, to actual detection of current surface changes (e.g. gullies & dune movement). For Venus, metre-scale imagery will enable study of Aeolian features and dunes (only two dune fields have been unambiguously identified to date on Venus); will enable more accurate stratigraphy and visibility of layering; will constrain the morphology of tesserae enough that their stress history and structural properties can be constrained; will enabled detailed study of styles of volcanism by enabling detailed mapping of volcanic vents and lava flows; and will enable direct search for surface changes due to volcanic activity and Aeolian activity. Metre-scale resolution would even enable search for changes in rotation rate due to surface-atmosphere momentum exchange, which could constrain internal structure (Karatekin et al., 2011).

The revolutions in radar performance go beyond just spatial resolution. Differential InSAR allows surface change detection at centimetre scale. This technique has been used to show surface deformations after earthquakes and volcanic eruptions on Earth

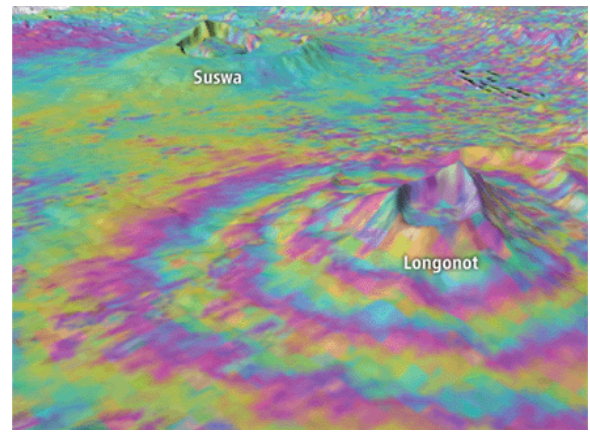

*Differential InSAR revealed altimetry changes in this volcano in Kenya which previously had been thought to be dormant. [Sparks et al., 2012; observations from Envisat ERS-1 & ERS-2].*

(see figure); similar results on Venus could provide dramatic evidence of current volcanic or tectonic activity.

The maturity of InSAR studies on Earth is such that data returned from Venus can confidently be understood within a solid theoretical framework developed from coupled terrestrial InSAR data and ground-truth observations, given the absence of ground-truth data on Venus. Note that the radar resolutions may be high enough to permit identification of Venera and Pioneer Venus landers, which serves as some ground truth even before a new generation of landers is taken into account. Radar mapping with different polarisation states constrains surface roughness and dielectric properties; mapping at different look angles and different wavelengths will provide further new constraints on surface properties.

The proximity of Venus to Earth, the relatively calm (if extreme) surface conditions and lack of water, the absence of a large satellite and its moderately well-known geoid and topography, all help to ease the technical demands on the mission. Europe is the world leader in radar systems and could, with minimal cost and development, adapt a GMES Sentinel-1 or NovaSAR-S modular array antenna for use at Venus, providing higher-resolution imagery, topography, geoid and interferometric change data that will revolutionise our understanding of surface and interior processes.

In addition to radar techniques, the surface can also be observed by exploiting near-infrared spectral window regions at wavelengths of 0.8 – 2.5 μm. On the nightside of Venus, thermal emission from the surface escapes to space in some of these spectral windows, allowing mapping of surface thermal emissivity, as demonstrated by VEx/VIRTIS [Mueller et al., 2008]. A new instrument optimised for this observation could map mineralogy and also monitor the surface for volcanic activity [Ghail et al., 2012].

**Case for Venus in situ geological investigations:**

The Venera and Vega missions returned data about the composition of Venus surface materials, but their accuracy is not sufficient to permit confident interpretation. The Venera and VEGA analyses of major elements (by XRF) did not return abundances of Na, and their data on Mg and Al are little more than detections at the 2σ level. Their analyses for K, U, and Th (by gamma rays) are imprecise, except for one (Venera 8) with extremely high K contents (~4% $K_2O$) and one (Venera 9) with a non-chondritic U/Th abundance ratio. The landers did not return data on other critical trace and minor elements, like Cr and Ni. In addition, the Venera and VEGA landers sampled only materials from the Venus lowlands – they did not target sites in any of the highland areas, the coronae, tesserae, nor the unique plateau construct of Ishtar Terra. Currently available instruments could provide much more precise analyses for major and minor elements, even within the engineering constraints of Venera-like landers.

A new generation of geologic instrumentation should be brought to the surface of Venus, including Raman/LIBS and XRF/XRD; this would would allow mineralogical, as well as merely elemental, composition. Such precise analyses would be welcome for basalts of Venus' lowland plains, but would be especially desirable for the highland tesserae and for Ishtar Terra. The tesserae may well represent ancient crust that predates the most recent volcanic resurfacing event and so provide a geochemical look into Venus' distant past. Ishtar Terra, too, may be composed (at least in part) of granitic rocks like Earth's continental crust, which required abundant water to form. Coronae samples will reveal how magmatic systems evolve on Venus in the absence of water but possibly in the presence of $CO_2$, $SO_2$ or other volatiles. Surface geological analysis would benefit from high temperature drilling/coring and sample processing capabilities, although further investigation will be needed to assess the extent to which this can be within the scope of an L-class mission opportunity. Long-lived seismological stations would also be valuable in the longer term but they too are considered outside the scope o the current call.

**Descent imaging** has not yet been performed by any Venus lander. Descent imaging of any landing site would be useful, particularly so for the tessera highlands where it would reveal the morphology of the highland surfaces and yield clues as to what weathering processes have been at work in these regions. Multi-wavelength imaging in near-infrared wavelengths would yield compositional information to provide further constraints on surface processes. In particular, descent imaging can establish whether near-surface weathering or real compositional differences are the root cause for near-IR emissivity variations seen from orbit, (Helbert et al. 2008) providing important ground truth for these orbital observations.

**Profiles of atmospheric composition in the near-surface atmosphere** would reveal which chemical cycles are responsible for maintaining the enormously high carbon dioxide concentrations in the atmosphere, and would also reveal details about surface-atmosphere exchanges of volatiles. Several mechanisms have been invoked for buffering the observed abundance of carbon dioxide, including the carbonate (Fegley & Treimann 1992) or pyrite-magnetite (Hashimoto & Abe, 1997) buffer hypotheses. Measurements of near-surface abundances and vertical gradients of trace gases, in particular $SO_2$, $H_2O$, CO and OCS, would enable discrimination between different hypotheses. Correlating these data with lander and orbiter data will reveal how important and widespread the sources and sinks of these species might be.

## 2.2 Planetary evolution as revealed by isotope geochemistry

Geology is a powerful witness to history, but it does not provide answers about evolution in the time before the formation of the oldest rocks, which on Venus were formed only about a billion years ago. For constraints on earlier evolution, we must turn to isotope geochemistry.

**Radiogenic noble gas isotopes** provide information about the degassing history of the planet. $^{40}$Ar, produced from the decay of long-lived $^{40}$K, has been continuously accumulated over >4 billion years and so its current abundance constrains the degree of volcanic/tectonic resurfacing throughout history. $^{129}$Xe and $^{130}$Xe are produced from the now extinct $^{129}$I and $^{244}$Pu respectively within the first 100Ma of the solar system's history. The depletion of these isotopes in the atmospheres of Mars and Earth reveal that these two planets underwent a vigorous early degassing and blow-off, although the mechanisms of this blow-off and of subsequent deliveries of materials from comets and meteorites vary according to different scenarios. Neither the bulk Xe abundance nor the abundances of its eight isotopes have ever been measured at Venus. Measurements of these abundances would provide entirely new constraints on early degassing history. $^{4}$He, produced in the mantle from long-lived U and Th decay, has an atmospheric lifetime of only a few hundred million years before it is lost by escape to space. Therefore its current atmospheric abundance provides constraints on recent outgassing and escape rates within the last $10^8 – 10^9$ years.

**Non-radiogenic noble gas isotopes** provide information about acquisition and loss of planet-forming material and volatiles. Venus is less depleted in Neon and Argon isotopes than are Earth and Mars, but its Xe and Kr isotopic abundances are still unknown (Xe isotopic abundances have not yet been measured, and past measurements of Kr abundance vary by an order of magnitude, providing little useful constraint). The significant fractionation of xenon on Earth and Mars can be attributed to massive blowoffs of the initial atmospheres in the period after the radiogenic creation of xenon from its parent elements, ~50-80 Myr after planet formation. However, it could also be that the fractionation is reflecting that of a source material delivered late in planetary formation, perhaps from very cold comets. If Venus has the same xenon fractionation pattern as Earth and Mars, this would support the idea that a common source of fractionated xenon material was delivered to all three planets, and weaken the case that these reflect large blowoffs. If we see a pattern with less Xe fractionation then that would support the blow-off theory for Earth and Mars. Considered together with other non-radiogenic isotopic abundances, this allows determination of the relative importance of EUV, impact-related or other early loss processes. These measurements would also allow tighter constraints on the how much of the gas inventory originated from the original accretion disk, how much came from the solar wind, and how much came later from planetesimals and comets. Late impacts such as the Earth's moon-forming impact also have an effect on noble gas isotopic ratios so can also be constrained through these measurements.

**Light element isotopic ratios**, namely H, C, O N etc, provide further insights into the origin and subsequent histories of planetary atmospheres. Measurement of $^{20}Ne/^{21}Ne/^{22}Ne$ and/or of $^{16}O/^{17}O/^{18}O$ would enable determination of whether Earth, Venus and Mars came from the same or from different parts of the protoplanetary nebula, i.e. are they are truly sibling planets of common origin. Venus' enhanced Deuterium to hydrogen ratio, 150 times greater than that found on Earth, suggests that hydrogen escape has played an important role in removing water from the atmosphere of Venus, removing more than a terrestrial ocean's worth of water during the first few hundred million years of the planet's evolution (Gillmann et al., 2009). $^{14}N/^{15}N$ ratios have been found to vary considerably in the solar system, with the solar wind, comets,

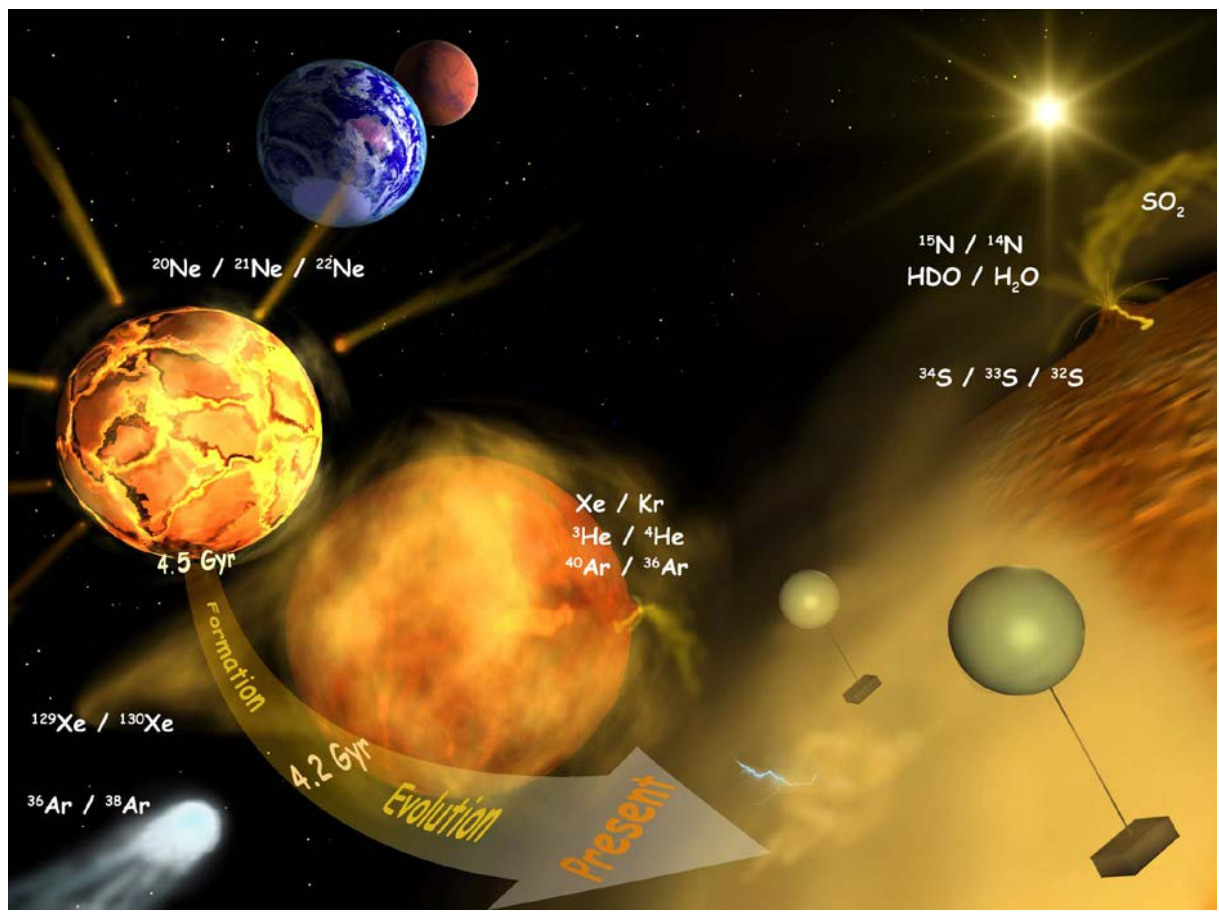

*Isotopic ratios provide keys to constrain planetary origins (Ne isotopes), early atmospheric loss processes (non-radiogenic Xe, Kr, Ar), late impact scenarios, recent mantle outgassing and late resurfacing, and escape of water. Modified from Baines et al., 2007.*

and meteorites all exhibiting isotopic ratios different from the terrestrial values; it also is affected by preferential escape of $^{14}$N. Measurement of the nitrogen isotopic ratio therefore helps establish whether the gas source was primarily meteoritic or cometary, and constrains history of escape rates.

Taken together, measurements of these isotopic abundances on Venus, Earth and Mars are needed to provide a consistent picture of the formation and evolution of these planets and their atmospheres, and in particular the history of water on Venus. Early Venus would have had an atmosphere rich in carbon dioxide and water vapour, like that of Hadean Earth. Hydrodynamic escape from this early steam atmosphere would have been rapid – but would it have been rapid enough to lose all of Venus' water before the planet had cooled enough to allow water to condense? If Venus did have a liquid water ocean, how long did this era persist before the runaway greenhouse warming 'ran away', with the oceans evaporating and the resulting water vapour being lost to space? The nature of early escape processes is as yet too poorly constrained to answer these questions. An early Venus with a liquid water ocean would have arguably have been more Earthlike than was early Mars and could have taken steps towards development of life. A habitable phase for early Venus would have important consequences for our understanding of astrobiology and the habitable zones of exoplanets.

More detailed treatments of Venus isotope geochemistry goals and interpretation cna be found in Chassefière et al., 2012 and Baines et al., 2007.

## 2.3 Atmospheric Science (Dynamics, Chemistry & clouds, structure & radiative balance)

The terrestrial planets today have very different climates. Study of the fundamental processes at work on these three planets will lead to a deeper understanding of how atmospheres work, and of climates evolve.

**Dynamics and thermal structure**

One of the crucial factors determining planetary habitability is the redistribution of heat around the planet. The solid planet of Venus rotates only once every 243 days but the atmosphere above exhibits strong super-rotation, circling the planet some 40-50 times faster than the solid planet below. General Circulation Models are now able to reproduce super-rotation, but are very sensitive not only to model parameters but also to the details of how those models operate. Efforts to improve modelling of the Venus atmosphere have led to improvements in how Earth handle details like conservation of angular momentum and temperature dependent specific heat capacities [Bengtsson et al., 2013]. Exchanges of momentum between surface and atmosphere, an important boundary condition for the atmospheric circulation, depend sensitively on the thermal structure in the lowest 10 km of the atmosphere. Many probes experienced instrument failure at these high temperatures so the atmospheric structure in the lowest parts of the atmosphere is not well known. The atmospheric circulation is driven by solar absorption in the upper cloud; most of this energy absorption is by an as yet unidentified UV absorber, which is spatially and temporally variable. Efforts to identify this UV absorber by remote sounding have failed, so in situ identification will be necessary.

Knowledge of the wind fields on Venus is currently achieved by tracking cloud features or by tracking descent probes. Tracking by cloud features returns information about winds at altitudes from 48 to 70 km altitude, depending on the wavelength used. However, it is not clear at what altitude to assume that the derived cloud vectors apply; the formation mechanism for the observed contrasts is not known so it cannot be ascertained to what extent the derived velocity vectors represent true air motion rather than the product of, for example, wave activity. Furthermore, cloud tracking on the day- and night-sides of Venus is accomplished using different wavelengths, referring to different altitudes, so global averages of wind fields are not achievable by wind tracking, frustrating attempts to understand global circulation.

Direct measurement of mesospheric wind velocities (or at least their line-of-sight components) from orbit can be achieved by using Doppler sub-millimetre observations, Doppler LIDAR or other such instruments, and this would help to constrain circulation models. Tracking of descent probes will yield direct measurement of the vertical profile of horizontal winds in the deep atmosphere, which will provide important constraints on the mechanisms of super-rotation. Balloon elements are ideal for measuring vertical wind speeds but also provide information about wave and tidal activity at constant altitude.

**Chemistry and clouds**

Venus has an enormous atmosphere with many complex chemical processes at work. Processes occurring near the surface, where carbon dioxide becomes supercritical and many metals would melt, are very different from those at the mesopause where temperatures can be below -150°C, colder than any found on Earth. A diversity of observations is clearly required to understand this diversity of environments. While chemical processes in the mesosphere and lower thermosphere are now being studied by Venus Express, processes in the clouds and below are very difficult to sound from orbit and require in situ investigation.

The dominant chemical cycles at work in Venus's clouds are those linking the sulphuric acid and sulphur dioxide: Sulphuric acid is photochemically produced at cloud-tops, has a net downwards transport through the clouds, and then evaporates and then thermal dissociates below the clouds; this is then balanced by net upwards transport through the clouds of its stable chemical precursors ($SO_2$ and $H_2O$). Infrared remote sensing observations of Venus can be matched by assuming a cloud composition entirely of sulphuric acid mixed with water, but Vega descent probe XRF measurements found also tantalising evidence of P, Cl and even Fe in the cloud particles [Andreichikov et al., 1987]. If confirmed these measurements would provide important clues as to exchanges with the surface: are these elements associated with volcanic, Aeolian or other processes? An in situ chemical laboratory floating in the clouds, or multiple descent probes, would be needed to address these measurement goals.

As to lower atmosphere chemistry, it is poorly understood because the rapidly falling descent probes did not have time to ingest and fully analyse many atmospheric samples during their brief descent. Sub-cloud hazes were detected by several probes but their composition is unknown. Ground- and space-based observations in near-infrared window regions permit remote sounding of only a few major gases, but many minor species which may play important catalytic or intermediate roles in chemical cycles cannot be probed remotely. Spatial variation of volcanic gases would be a direct way of finding active volcanism on Venus but would be very difficult to achieve. However, as mentioned above, near surface abundances and their vertical profiles will permit determination of whether there are active surface-atmosphere exchanges taking place and of what surface reactions are buffering the atmospheric composition.

**Thermal structure and radiative fluxes**

The cloud layer of Venus is highly reflective so Venus presently absorbs less power from the sun than does the Earth. Its high surface temperature is instead caused by its enormous greenhouse warming effect, caused by carbon dioxide, water vapour and other gases. Its clouds, too, have a net warming effect because they prevent thermal fluxes from escaping the deep atmosphere.

1-D radiative and radiative-convective models for the determination of climate are suddenly widespread as researchers worldwide attempt to determine the likely climate of exoplanets. Venus offers a proving ground for these models much closer to home, one where the conditions are much better known than on exoplanets. Radiative transfer calculations on Venus are difficult: uncertainties in the radiative transfer properties of carbon dioxide at high temperatures and pressures are the main unknown, particularly in the middle- and far- infrared where there are no spectral window regions to allow empirical correction. As on Earth, clouds play an important role, reflecting away sunlight but also trapping upwelling infrared radiation. The state-of-the art Venus radiative balance are still only 1-D models representing an average over the whole planet. However, we now know that the clouds are very variable; the vertically integrated optical thickness (as measured at 0.63 µm) can vary by up to 100% [Barstow et al 2012] and the vertical structure of clouds varies strongly with latitude. In-situ measurements of cloud properties with co-located radiative flux measurements are needed to determine the diversity of cloud effects on the global radiative balance.

## 3 Mission elements

**A satellite in low, near-circular polar orbit** is required for radar mapping, and for LIDAR and Doppler sub-mm measurement of wind speeds. Wide angle cameras like the MRO/MARCI camera can be used to obtain continuous imaging coverage of UV cloud-top features. Recent examples of proposed low circular Venus orbiters include the Envision radar mapper [Ghail et al., 2012], the RAVEN radar mapper [Sharpton et al., AGU 2009], VERITAS radar mapper [Hensley, Smrekar et al., AGU 2012], MuSAR radar mapper [Blumberg, Mackwell et al., URSI 2011], and the Vesper sub-mm sounding orbiter [Allen et al., DPS 1998].

A **satellite in a highly elliptical orbit** can provide synoptic views of an entire hemisphere at once; One example of this is Venus Express, whose polar apocentre allows it to study the vortex circulation of the South polar region; a second example is the nearly equatorial 40-hour orbit of Japan's Akatsuki orbiter, which allows it to dwell over low-latitude cloud features for tens of hours at a time. The large range of altitudes covered is also useful for in situ studies of thermosphere and ionosphere and thus for studies of solar wind interaction and escape.

**Balloons** are ideally suited for exploring Venus because they can operate at altitudes where pressures and temperatures are far more benign than at the surface. Deployment of two small balloons at 55 km altitude, in the heart of the main convective cloud layer, was successfully demonstrated by the Soviet VeGa mission in 1984. At this altitude, the ambient temperature is a comfortable 20° C and the pressure is 0.5 atm. The main environmental hazard is the concentrated sulphuric acid which makes up the cloud particles; however, effects can be mitigated by choosing appropriate materials for external surfaces. Balloons at this altitude can take advantage of the fast super-rotating winds which will carry the balloon all the way around the planet in a week or less (depending on latitude and altitude). Horizontal propulsion (with motors) is not advised because of power requirements and the difficulty of countering the fast (250 km/h) zonal windspeed. A cloud-level balloon is an ideal platform for studying interlinked dynamical chemical and radiative cloud-level processes. It also offers a thermally stable long-lived platform from which measurements of noble gas abundances and isotopic ratios can be carefully carried out and repeated if necessary (in contrast to a descent probe, which offers one chance for making this measurement, in a rapidly changing thermal environment).

Balloons can be used to explore a range of altitudes. Operation in the convectively stable upper clouds, above 63 km, would be optimal for identification of the UV absorber, but the low atmospheric density leads to a relatively small mass fraction for scientific payload. Operation below the main cloud deck at 40 km, has been proposed by Japanese researchers, with a primary goal of establishing wind fields below the clouds. Balloons can also be used to image the surface, if they are within the lowest 1-10 km of the atmosphere, but high temperatures here require exotic designs such as metallic bellows which are beyond the scope of this paper. An intriguing possibility for revealing winds in the lower atmosphere is to use passive balloons, reflective to radio waves, which could be tracked by radar – this possibility should be studied further if a radar + entry probes architecture were to be studied further.

**Descent probes** provide vertical profiles of composition, radiation, chemical composition as a function of altitude, and enable access to the surface. Science goals for a descent probe include: cloud-level composition and microphysical processes; near-surface composition, winds, and temperature structure; surface composition and imaging; and noble gas abundances and isotopic ratios. If the scientific focus of the probe is measurements at cloud level then a parachute may be deployed during the initial part of the entry phase in order to slow the rate of descent during the clouds. This was carried out, for example, by the Pioneer Venus Large Probe. Alternatively, if the main focus of the probe is measurements in the lower atmosphere or surface then the probe may dispense with a parachute completely. Descent imagery of impact/landing site at visible wavelengths will be invaluable for contextualising surface results; Rayleigh scattering limits the altitudes from which useful surface imagery can be obtained to ~1 km if imaging in visible wavelengths, or to 10 km if imaging at 1 μm wavelength [Moroz, PSS 2002].

**Surface elements** – including static landers and rovers – must cope with the harsh conditions of ~450 °C at the surface of Venus. Rovers and landers which require sustained surface operations would require nuclear power due to the low levels of sunlight reaching the surface, and are considered beyond the scope of this proposal. However, a surface element using passive thermal control – relying on thermal inertia and thermal insulation to keep a central electronics compartment cool – allows operation times of hours or even days [see e.g. Venera-D mission, Vorontsov et al., Solar System Research 2011]. The principal science payload of a surface element would include surface imagers and non-contact mineralogical sensors such as Gamma spectrometer with Neutron activation, capable of measuring elemental abundances of U, Th, K, Si, Fe, Al, Ca, Mg, Mn, Cl (Li, Mitrofanov et al., EPSC 2010) and/or Raman/LIBS (Clegg et al., LPSC 2011). Inclusion of surface sample ingestion via a drill/grinder/scoop would allow further analysis techniques (e.g. mass spectroscopy; X-ray fluorescence (XRF) spectroscopy) but would require significant technology development and verification. Gamma- and XRF spectroscopy have been performed on Venera and Vega landers, but modern equivalents of these instruments would provide much improved accuracy; also, repeating the composition analyses at a tessera region (not before sampled) would reveal whether these tessera regions are chemically differentiated from the lava plains where previous analyses have been conducted.

# 4  A strawman mission architecture

A Large mission to Venus should include both orbital and in situ science measurements. One possible strawman mission concept which would could address this theme would be a combination of **an orbiter, a cloud-level balloon platform, and (optionally) a Russian descent probe**. As a strawman payload, we suggest the balloon element be modelled on the 2010 EVE M3 proposal [Wilson et al., 2012]. The radar orbiter may be based on a reuse of ESA's GMES Sentinel-1 InSAR technology, whose application at Venus was first described in the 2010 Envision M3 proposal [Ghail et al., 2012]. Finally, the landing probe envisaged is based on the lander component of the Venera-D mission [Vorontsov et al., 2011]

It is very important to have multiple mission elements working together simultaneously at Venus. An orbiter is necessary both to increase vastly the volume of data returned from the in situ elements, but also to place those in situ measurements into atmospheric and geological context. The in situ measurements are required to measure parameters, like noble gas abundances and surface mineralogy, which cannot be determined from orbit. The whole mission is greater than the sum of its parts. This has been amply demonstrated by the constellation of missions at Mars, and by the Cassini/Huygens collaboration at Titan.

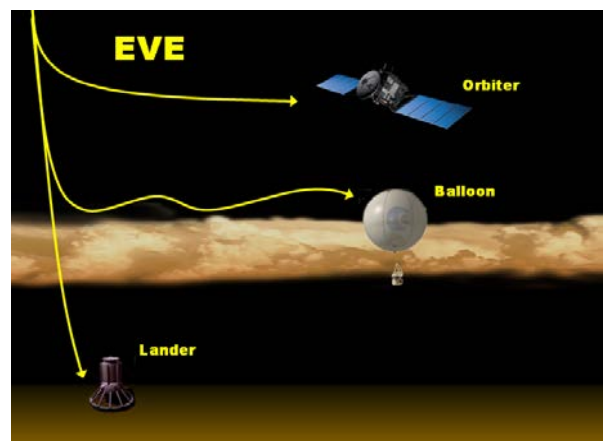

Not all of these mission elements need be provided by ESA; There is ample scope for international co-operation in creating this mission architecture. In particular, Russia has unequalled heritage in providing Venus descent probes from its Venera and Vega descent probes, and will gain new heritage from its Venera-D lander, planned for the coming decade. A range of mission proposals have been developed in the USA for orbiters, balloons and descent probes, from Discovery-class to Flagship-class, many of which could form parts of a joint NASA-ESA exploration programme should a high-level agreement be reached. Japan has an active Venus research community, with its Akatsuki (Venus Climate Orbiter) spacecraft still in flight, and has been developing prototypes for a Venus sub-cloud balloon [Fujita et al., IPPW 2012]. After its recent successful launches of radar satellites RISAT-1 and RISAT-2, ISRO has also investigated possibilities of an Indian Venus mission [Anurup et al., 2012], by 2030 this may be a real possibility. Israel's TECSAR satellites enable 1-m scale radar mapping with a 300 kg satellite, in the frame of future ESA-Israel agreements, collaborations on Venus radar could be fruitful. In this time frame collaborations with China are also feasible.

These could be launched as a stack on a single launcher, or it may prove convenient to use separate launchers, for example in order to insert the orbiter into a low circular orbit before the arrival of the in situ elements for optimal data relay and context remote sounding for the in situ measurements.

This scenario, Orbiter + balloon + Descent Probe, was proposed to ESA in 2007 in response to the M1/M2 mission Call for Ideas as a joint European Russian mission, with a European-led orbiter and balloon, and a Russian descent probe. For EVE 2007, the entire mission was to be launched on a single Soyuz launch, however subsequent studies revealed that this scenario was not consistent with a single Soyuz launcher, and would be more consistent with an L-class rather than an M-class opportunity.

# 5 Technology developments needed

Much of the technology required for a Large Venus mission already exists at a high Technology Readiness Level, but further technology development both for spacecraft technologies and for science payload technologies would be useful to maximise science return and de-risk mission aspects.

Many of the mission-enabling technologies required for Venus exploration are shared with other targets. Aerobraking/aerocapture would improve Δv and mass budgets for Venus orbiters. Further improvement in deep space communications, including development of Ka-band or optical communications, would provide increased data return from orbiters which would be particularly useful for the large volumes of data generated by high-resolution radar instrumentation at Venus – 100 Mbps or better would be enable the maximum scientific return from a radar orbiter. High speed entry modelling, thermal protection systems and parachute development will all be useful for Venus entry probes. Nuclear power systems are not necessary for the strawman Venus mission described here, but would enable longer lifetimes and increased nightside operations for the balloon element of the mission, and will be necessary for long-lived surface stations (in the even more distant future) due to the scarcity of sunlight reaching the Venus surface and long night-time duration.

Two technology areas specific to Venus exploration are balloon technology, and high-temperature components. The balloons proposed in this strawman mission are helium superpressure balloons, which are designed to float at constant altitude. Although ESA has little familiarity with this technique, thousands of helium superpressure balloons have been launched on Earth, and two were successfuly deployed on Venus in 1985 as part of the Russian VeGa programme, so this is a very mature technology. Air-launching of balloons – deploying them from a probe descending under parachute – was achieved by the Russian VeGa balloons, was demonstrated by CNES in VeGa development programmes, and demonstrated recently by JPL engineers in their own Venus balloon test programme; nevertheless, a new demonstration programme in Europe would be required to obtain recent European experience for this technology. Balloon envelope design and sulphuric acid resistance verification would also be valuable to conduct in Europe. Feasibility studies on other forms of aerial mobility, including phase change fluid balloons, <5 kg microprobes and fixed wing aircraft, would also be useful for expanding the possibilities of long-term future exploration programmes. Compact X-band phased array antennae are well suited for mobile atmospheric platforms like balloons (or indeed rovers), investment in these systems would be valuable.

High temperature technologies are only needed if Europe is to provide mission elements or payloads which need to operate in the lower atmosphere, below 40 km altitude. The descent probe proposed in the present strawman mission proposal is a classical Venera-type design with all electronics and most sensors inside a layer thermal insulation in a central compartment with high thermal inertia. This kind of passive thermal design permits operation on the surface for hours or even days. The more components can be placed on the outside of the probe, the longer the mission lifetime can be extended. Development programmes in high-temperature electronic components like amplifiers for telecommunications systems would therefore be valuable.

Further investment in scientific payload development is also needed to get the most science return from a large Venus mission. The radar is a critical element of the mission and further optimisation with respect to that proposed in the Envision proposal is still possible. Thanks to technology developments in the last ten years (including ESA's Sentinel-1A mission, Israel's TECSAR mission and the UK's NovaSar-S developments), metre-scale InSAR mapping has become not only possible but also affordable and deployable in small spacecraft. Further development would be valuable to adapt these systems for Venus, including further work on surface height change detection through differential interferometric SAR. Investment in optical and sub-mm heterodyne receivers would hasten the development of instruments capable of measuring mesospheric wind velocities using Doppler techniques.

For atmospheric in situ measurements, a key instrument is a mass spectrometer with getters and cryotraps to isolate and precisely measure the noble gas and light element isotope abundances. These technologies have been developed for Mars (e.g. MSL/SAM, ExoMars/PALOMA proposal), but further development to maximise the precision of the measurements and to optimise the development for the thermal environment of Venus balloons and descent probes will be needed. In situ GC+MS characterisation atmospheric chemistry is another mature field, but further development in particular of an aerosol collector system to allow detailed characterisation of cloud particle composition would be valuable.

Specific payload developments for a landing probe should also include surface characterisation instruments – gamma ray and neutron spectrometers, XRF/XRD, Raman instruments for mineralogical identification. High-temperature drilling and sample ingestion systems are not currently proposed for the Venera-D lander included in the strawman mission, but some feasibility studies in this area would be a useful investment. Finally, high temperature chemical, meteorological and seismological sensors using silicon carbide semiconductor technology would usefully complement a lander's payload.

In summary, most of the technologies needed for a Large class Venus mission already have high levels of heritage either from Earth or from other planetary missions, but a well-targeted development programme would de-risk mission elements and maximise the science return. Many of the development areas identified would also benefit other space missions and have spin-out potential on Earth.

# 6 Conclusions

*As we become aware of Earth's changing climate, and as we discover terrestrial planets in other solar systems, we gain ever more reasons to study the Earth's nearest neighbour and closest sibling.*

*For the scientific and programmatic reasons outlined in this document, Venus is a compelling target for exploration. The science themes important for Venus research – comparative planetology and planetary evolution – are common to all of planetary and exoplanetary science, and many of the instrumentation required – in situ mass spectrometry, radar and atmospheric remote sensing – are found in mission proposals for many other solar system targets. Venus is close to the Earth, which leads to a short cruise phase (typically only 5 months) and high data rates, and also close to the sun, resulting in plentiful solar power with modest solar arrays. These factors ensure that it will be a highly attractive target for emerging space nations to send missions, so is an ideal arena for inter-agency collaboration.*

*Venus is thus an excellent proving ground for international collaboration on large space missions; an excellent proving ground for new instrumentation before it is sent on long journeys to the distant reaches of the solar system; an excellent proving ground for techniques of analysis of exoplanets; an excellent proving ground for fundamental understanding of geophysical processes of terrestrial planets; an indispensable part of our quest to understand the evolution of Earthlike planets. For all these reasons, a Venus mission would be a strong candidate for an ESA Large-class mission in the coming decades and we therefore propose science themes of planetary evolution and comparative planetology for the L2/L3 opportunities.*